# Intentional Aliasing Method to Improve Sub-Nyquist Sampling System

Jehyuk Jang, Sanghun Im, and Heung-No Lee*, *Senior Member, IEEE*

*Abstract*—A modulated wideband converter (MWC) has been introduced as a sub-Nyquist sampler that exploits a set of fast alternating pseudo random (PR) signals. Through parallel analog channels, an MWC compresses a multiband spectrum by mixing it with PR signals in the time domain, and acquires its sub-Nyquist samples. Previously, the ratio of compression was fully dependent on the specifications of PR signals. That is, to further reduce the sampling rate without information loss, faster and longer-period PR signals were needed. However, the implementation of such PR signal generators results in high power consumption and large fabrication area. In this paper, we propose a novel aliased modulated wideband converter (AMWC), which can further reduce the sampling rate of MWC with fixed PR signals. The main idea is to induce intentional signal aliasing at the analog-to-digital converter (ADC). In addition to the first spectral compression by the signal mixer, the intentional aliasing compresses the mixed spectrum once again. We demonstrate that AMWC reduces the number of analog channels and the rate of ADC for lossless sub-Nyquist sampling without needing to upgrade the speed or the period of PR signals. Conversely, for a given fixed number of analog channels and sampling rate, AMWC improves the performance of signal reconstruction.

## I. INTRODUCTION

Applications of electronic warfare (EW) systems, electronic intelligence (ELINT) systems, or cognitive radios are demanding the observation of a multiband signal, i.e., a collection of multiple narrow-band signals, each with different center frequencies, scattered across a wide frequency range up to tens of gigahertz (GHz). The Nyquist sampling rate is twice the maximum frequency of the wide range. When a multiband signal is *sparse*, i.e. consists of a few narrow bands, the signal can be sampled without information loss at a sub-Nyquist rate far less than the Nyquist rate. The theoretical lower limit of the rate required for lossless sub-Nyquist sampling is the sum of the bandwidths, known as the Landau rate, when the spectral locations of all the narrow-band signals are known [1]. When spectral locations are unknown, the lower limit is doubled [2].

The modulated wideband converter (MWC) proposed by Mishali *et al.* [3] is a lossless sub-Nyquist sampler that aims at achieving the theoretical lower limit of sampling rate.

Similar to other sub-Nyquist samplers proposed in [4]–[6], MWC exploits pseudo-random (PR) signals, which periodically output pulsed signals. MWC has multiple analog channels, each of which consists of a PR signal generator, signal mixer, low-pass filter (LPF) for anti-aliasing, and low-rate analog-to-digital converter (ADC) in sequence. The system compresses a multiband spectrum through the mixing and LPF procedures, following which it samples at a sub-Nyquist rate. The reconstruction of the input multiband spectrum is guaranteed under some conditions of the compressed sensing (CS) theory [7]–[11]. With the help of CS reconstruction algorithms in [2], [12] developed for the MWCs, it has been proved that an MWC can achieve the theoretical lower limit of the lossless sub-Nyquist sampling rate.

However, to achieve the lower limit of the lossless sub-Nyquist sampling rate, the previously proposed MWC by Mishali *et al.* relied on a high-end PR signal generator, since it was the only spectral compressor. The ratio of spectral compression was fully dependent on the *oscillation speed* and *length of the pulsed patterns* within a single period of the PR signals. Specifically, to improve the compression ratio for a sparser multiband signal, PR signals with a greater pattern length were required. In addition, the oscillation speed should be faster than the Nyquist rate for a lossless compression. Unfortunately, increasing the pattern length of a PR signal generator with tens of GHz-range switching speed leads to difficult research problems in the field of chip engineering, such as high power consumption and large fabrication area due to the high chip speed [13], [14], which hinder the commercial availability of such a PR signal generator chip.

Recently, efforts to reduce the rate for lossless sub-Nyquist sampling with MWC closer to the theoretical lower limit without upgrading the PR signal generators have been made in [15], [16]. In [15], the authors proposed a method that channelizes the multiband spectrum into few orthogonal subbands before mixing with the PR signals. Since the channelized signals have a lower Nyquist rate than the original input, for a given oscillation speed and pattern length of PR signals, the method achieves a higher ratio of spectral compression. Although the method led to a further reduction of the lossless sub-Nyquist sampling rate, it requires additional hardware resources for the channelization, such as band-pass filters, local oscillators, and a greater number of independent PR signal generators proportional to the number of subbands. In [16], a method similar to that proposed in [15] was presented, in which the input signal was divided into in-phase (I) and quadrature (Q) channels before mixing it with PR signals. The lossless sub-Nyquist sampling rate can be reduced by the same principle as in [15], although the authors did not mention this point. However, the system also required additional hardware resources for the I-Q division.

This work was supported by a grant-in-aid of HANWHA THALES.

J. Jang and the corresponding author Heung-No Lee are with the Department of Electrical Engineering and Computer Science, Gwangju Institute of Science and Technology, Gwangju 61005, South Korea (e-mail: jjh2014@gist.ac.kr; heungno@gist.ac.kr).

S. Im is with Hanwha Systems, Seongnam 13524, South Korea (e-mail: sh.im@hanwha.com)

In this paper, we propose an aliased MWC (`AMWC`), which reduces the lossless sub-Nyquist sampling rate for given practical PR signals. The main idea of `AMWC` is to break the anti-aliasing rule and induce *intentional aliasing* at the ADC of each spatial channel by setting the bandwidth of the prior LPF to be greater than the ADC sampling rate. In addition to the first spectral compression by the mixing and LPF procedures, this intentional aliasing leads to another spectral compression under a certain relation between the ADC sampling rate and bandwidth of the prior LPF. Through the two spectral compression procedures, the compression ratio is improved without faster or longer PR signals. Consequently, for a given and fixed PR signal generator, the lossless sub-Nyquist sampling rate of `AMWC` is closer to the lower limit than that of MWC.

The proposed `AMWC` achieves the same effect as in previous works [15], [16], i.e., reduction in the lossless sub-Nyquist sampling rate without upgrading the PR signal generators, and requires no additional hardware components. To our knowledge, `AMWC` is *novel* in that no study has thus far improved the sub-Nyquist sampling capability of MWC by improving the utilization efficiency of given hardware resources.

In [17], [18], variations of MWC similar to `AMWC` that include aliasing at the ADC have been investigated for analyzing *channel capacity*. Their main results indicate that suppressing non-active subbands before spectral compression minimizes the loss of information rate incurred by aliasing the noise spectrum. Interestingly, the authors of [18] introduced a rule for determining the sampling rate of each spatial channel similar to that of `AMWC` (see Section III-A for details). However, the rule was designed to make a fair comparison with other filterbank-based systems by flexibly controlling the bandwidth of subbands, rather than to exploit the aliasing at the ADC to reduce the lossless sub-Nyquist sampling rate. Additionally, according to our results, the rule in [18] is insufficient and aliasing at the ADC may lead to information loss.

Our main contribution is that the anti-aliasing rule of MWC is shown to be unnecessary for lossless sub-Nyquist sampling. We reveal a certain relationship between the ADC sampling rate and bandwidth of the prior LPF so that `AMWC` can avoid the loss of signal information during the additional spectral compression. We demonstrate that, for given oscillation speed and pattern length of PR signals, the sampling rate and analog channels of `AMWC` required for the reconstruction of a multiband signal are further reduced. For given sampling rate and number of analog channels, we show that the reconstruction performance of `AMWC` for a multiband signal with a given sparsity is improved.

Additionally, we show that the benefits from intentional aliasing can be further strengthened using a non-flat LPF. The non-flat frequency response of LPF results in a different input-output relationship for each frequency component of the sub-Nyquist samples of `AMWC`. Simulation results show that the reduction of lossless sub-Nyquist sampling rate is boosted when the filter response is samples of a random distribution as the input-output relationships of different frequency components become independent.

The remainder of this paper is organized as follows. In Section II, we briefly introduce MWC with the anti-aliasing rule and then define the goal of this paper. In Section III, we propose `AMWC` and derive its input-output relationship. The relationship between the sampling rate of ADC and bandwidth of LPF to avoid information loss is also provided. In Section IV, a revised input-output relationship of `AMWC` corresponding to the use of a non-ideal LPF is provided. Simulation results are provided in Section V. Section VI concludes the paper.

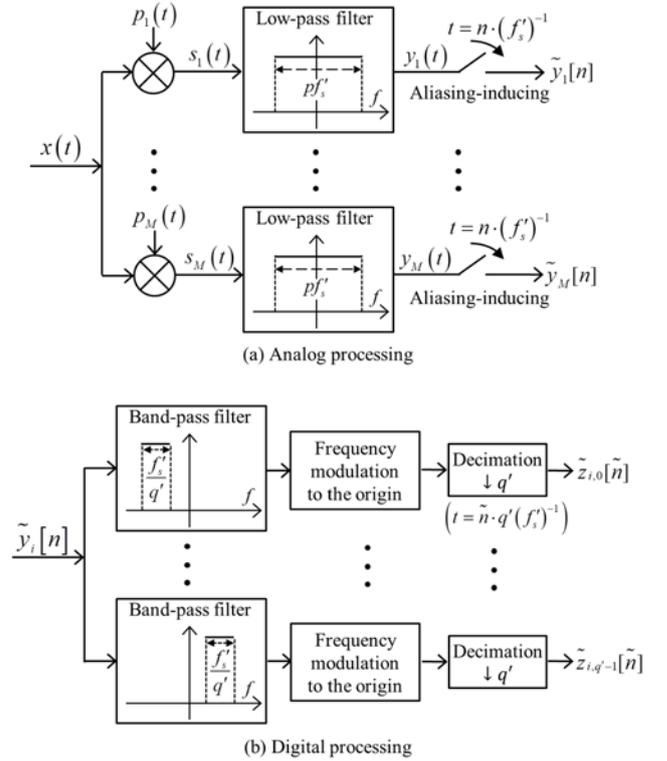

Fig. 1. Sampling system of `AMWC`. The system is equivalent to `cMWC` when $p=1$ and $q'=q$ . In `AMWC`, the sampling rate is $p$ -times lower than the filter bandwidth with $p>1$ to intentionally induce aliasing.

## II. Background and Problem Formulation

The modulated wideband converter (`MWC`) is a sub-Nyquist sampling system for multiband signals. A signal $x(t)$ is a multiband signal if its spectrum $X(f)$ is composed of $K_B$ disjoint continuous bands of maximum bandwidth $B$ [2], [3]. We assume that the maximum frequency of a target multiband signal does not exceed $f_{max}$ , i.e., $X(f)=0$ for $f \in \mathcal{F}_{NYQ}{}^c$ , where $\mathcal{F}_{NYQ} \triangleq [-f_{max}, f_{max})$ , and $\mathcal{F}_{NYQ}{}^c$ is the complementary set of $\mathcal{F}_{NYQ}$ . We denote the Nyquist rate by $f_{NYQ} \triangleq 2 f_{max}$ .

### A. System Constitution and Parameters

`MWC` consists of $M$ analog channels in parallel (see Fig. 1-(a)). Each channel consists of a PR signal generator, a mixer, an LPF, and an ADC in sequence. Each PR signal $p_i(t)$ for channel index $i$ is $T_p$ -periodic and outputs chips of an odd length $L$ within a single period $T_p$ . Each chip lasts for a chip duration $T_c = T_p L^{-1}$ . We denote the *chip speed* by $f_c \triangleq T_c{}^{-1}$ and the repetition rate of the PR signal by $f_p \triangleq T_p{}^{-1}$ . The LPF has a cut-off frequency $W_{LPF}/2$ , where $W_{LPF}$ denotes the bandwidth of the filter including the negative frequency. The LPF bandwidth is set to $W_{LPF} = qf_p$, where $q$ is the *channel-trading parameter*, an odd positive integer. Finally, we denote the sampling rate, which is equal at every channel, by $f_s$. The *total sampling rate* is the sum of sampling rates of all channels, defined by $f_{s,total} \triangleq Mf_s$.

MWC first compresses the input multiband spectrum using PR signals. After that, nonzero subbands of the multiband spectrum are recovered by CS recovery algorithms. For the successful CS recovery, all spectral components within the Nyquist range $\mathcal{F}_{NYQ}$ of each PR signal are needed to be independent, which requires a fast chip speed $f_c \geq f_{NYQ}$ [3]. Throughout this paper, we set $f_c = f_{NYQ}$.

### B. Conventional Modulated Wideband Converters

In the original paper [3] by Mishali *et al.*, for lossless sub-Nyquist sampling, the ADC followed the *anti-aliasing* rule, i.e., $f_s \geq W_{LPF}$. This conventional rule has sufficed for lossless sub-Nyquist sampling. We refer to MWC that follows the anti-aliasing rule as *conventional MWC* (cMWC).

The input-output relationship of cMWC is given in [3]. The input $x(t)$ at the $i$-th channel is first mixed with the $T_p$-periodic PR signal $p_i(t)$ that periodically outputs a sequence of $L$ mixing chips. By the periodicity, the Fourier transform (FT) of $p_i(t)$ is an impulse train. The FT of the mixed signal $s_i(t) = x(t)p_i(t)$ is the convolution $*$ of the two spectra:

$$
\begin{aligned}
S_i(f) &\triangleq \int_{-\infty}^{\infty} s_i(t) e^{-j2\pi ft} dt \\
&= P_i(f) * X(f) \\
&= \sum_{l=-\infty}^{\infty} c_{i,l} X(f - lf_p),
\end{aligned}
\tag{1}
$$

where $c_{i,l}$ for $l = -\infty, \cdots, \infty$ are the Fourier series coefficients of $p_i(t)$. The mixed signal $s_i(t)$ and $X(f - lf_p)$ in (1) are filtered by the LPF $H(f)$. We let $H(f) = 1$ for $f \in \mathcal{F}_{LPF}$, and otherwise, $H(f) = 0$, where $\mathcal{F}_{LPF} \triangleq [-W_{LPF}/2, W_{LPF}/2]$. Since $X(f)$ is band-limited by $\mathcal{F}_{NYQ}$, the infinite-order summation in (1) is reduced to a finite order as follows:

$$
\begin{aligned}
Y_i(f) &= S_i(f) H(f) \\
&= \sum_{l=-(L_0+q_0)}^{L_0+q_0} c_{i,l} X(f - lf_p), \text{ for } f \in \mathcal{F}_{LPF},
\end{aligned}
\tag{2}
$$

where $L_0$ is computed by $L_0 = (L-1)/2$ [3], and $q_0 \triangleq (q-1)/2$. Next, the ADC of rate $f_s = T_s^{-1}$ takes samples of $y_i(t)$, i.e., $y_i[n] = y_i(t)|_{t=nT_s}$. By the conventional anti-aliasing rule, we set $f_s = W_{LPF}$. Then, the discrete-time FT (DTFT) of $y_i[n]$ preserves the spectrum of (2).

In (2), every subband $X(f - lf_p)$ is spectrally correlated with nearby $q-1$ subbands, since the bandwidth $W_{LPF}$ is wider than the shifting interval $f_p$. To make them spectrally orthogonal, the samples $y_i[n]$ are modulated and low-pass filtered in parallel through $q$ digital channels by

$$
z_{i,s}[\tilde{n}] = \left[ \left( y_i[n] e^{-j2\pi s f_p T_s n} \right) * h_{f_p}[n] \right]_{n=\tilde{n}q}
\tag{3}
$$

for $s = -q_0, \cdots, q_0$, where $h_{f_p}[n]$ is a digital LPF with the cut-off frequency of $f_p/2$ and a flat passband response. The DTFT of (3) is

$$
Z_{i,s}\left( e^{j2\pi f q T_s} \right) = \sum_{l=-L_0}^{L_0} c_{i,l+s} X(f - lf_p) \text{ for } f \in \mathcal{F}_p,
\tag{4}
$$

where $\mathcal{F}_p \triangleq [-f_p/2, f_p/2]$. The subbands $X(f - lf_p)$ in (4) are spectrally orthogonal to each other, since the bandwidth equals the shifting interval. As $X(f)$ is a multiband signal, only a few subbands in (4) have nonzero values. If $f_p \geq B$, the upper bound on the sparsity $K$ of the subbands is $K \leq 2K_B$, since the uniform grid of interval $f_p$ splits each band into two pieces at most.

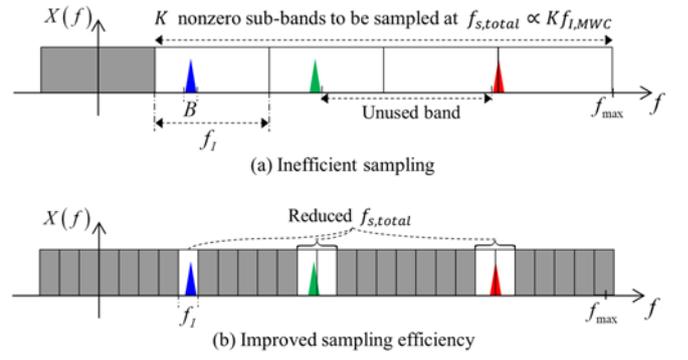

Fig. 2. Illustration of the *sampling efficiency* in the relation between the maximum bandwidth $B$ and splitting interval $f_l$. (a) When $f_l \gg B$, MWC wastes a portion of the total sampling rate because of the unused band in the nonzero subbands. (b) The regulated $f_l$ improves the sampling efficiency.

Consequently, each analog channel outputs $q$ different sequences, and therefore, cMWC obtains totally $Mq$ equations for input reconstruction. Depending on the number of equations, it was shown in [3] that the input spectrum can be perfectly reconstructed. Previously, to obtain more equations for a fixed number of channels $M$ and for a given specification $f_p$ for PR signal generation, cMWC has to rely on the increased sampling rate $f_s = qf_p$ by controlling the *channel-trading* parameter $q$. In this paper, we aim to show there is another way to obtain more equations and improve the input reconstruction performance, without the cost intensive ways of increasing the total sampling rate $f_{s,total} = Mf_s$ or reducing $f_p$, or both.

### C. Sampling Efficiency

In (4), MWC splits the input spectrum into many subbands along a uniform grid of a *splitting interval*, and it then takes samples of the weighted sum of subbands. We denote the splitting interval by $f_l$. Note that the splitting interval of cMWC $f_{l,cMWC}$ equals $f_p$. From the samples, a CS recovery algorithm (e.g., [11], [12], [19], [20]) finally recovers the $K$ nonzero subbands containing the split pieces of the $K_B$ multibands. Consequently, the total sampling rate is consumed to take samples of $K$ nonzero subbands of bandwidth $f_I$. This indicates that the total sampling rate required for lossless sampling by an MWC would be at least $f_{s,total} \geq 2Kf_I$, where the factor of 2 arises from the unknown supports of the nonzero subbands. In contrast, a result in [2] states that, for a general sub-Nyquist sampling system, the minimum requirement for lossless sampling of a multiband signal is $f_{s,total} \geq 2K_B B$, where $K_B B$ is the upper bound of the *actual spectral occupancy* of a multiband signal. That is, when $f_I$ is far greater than $B$, MWC consumes a portion of the total sampling rate inefficiently. Specifically, $f_I$ greater than $B$ yields a higher probability for the $K$ nonzero subbands to be comprised of unused bands, i.e., zeros. The inefficient use of total sampling rate is illustrated in Fig. 2.

Ideally, when the splitting interval $f_I$ becomes finer and closer to $B$ while satisfying $f_I \geq B$, the sampling efficiency is improved, as shown in Fig. 2. The efficiency is maximized when $Kf_I = K_B B$. Based on this observation, we define the *sampling efficiency* $\alpha$ of MWC as the ratio between the actual spectral occupancy of the multiband signal and the total bandwidth of the recovered subbands, i.e.,

$$\alpha \triangleq \frac{K_B B}{Kf_I}. \qquad (5)$$

Note that, by the definition of $K$, $\alpha \leq 1$ always holds.

In summary, improving $\alpha$ has two advantages. First, for the lossless sampling of a given multiband signal, it would reduce the required total sampling rate $f_{s,total}$ closer to the theoretical minimum requirement $f_{s,total} \geq 2K_B B$. By the definition, the higher $\alpha$ closer to 1 indicates that a portion of $f_{s,total}$ inefficiently consumed for taking samples of the unused bands in Fig. 2 is reduced. By the reduced $f_{s,total}$, the number of channels $M$ or the sampling rate $f_s$ of ADC at each channel is reduced. Secondly, for given and fixed $f_{s,total}$, we will show throughout the rest of paper that improving $\alpha$ yields more independent equations for signal reconstruction, and thus, more complex multiband signals with higher $K_B$ can be recovered perfectly.

### D. Limitations in Conventional MWC

For cMWC, the sampling efficiency depends entirely on the hardware capabilities of PR signal generators, which may result in severe implementation problems. The sampling efficiency of cMWC depends on the specifications of PR signal generators since $f_{I,cMWC}$ is fixed to $f_p$. By the definition, the only way to improve the sampling efficiency $\alpha_{cMWC}$ of cMWC has been to make the repetition rate $f_p$ of the PR signals closer to $B$. As discussed, the chip speed $f_c$ of PR signals should not be less than the Nyquist rate, i.e., $f_c \geq f_{NYQ}$. Thus, from the relation $f_p = f_c L^{-1}$, the chip length $L$ is the only free parameter to control $f_p$. Since $B$ is usually far smaller than $f_{NYQ}$, to fit $f_p$ closer to $B$, a very long $L$ is needed. However, in applications where $f_{NYQ}$ reaches tens of gigahertz, due to the extremely high chip speed $f_c$, implementing PR signal generators having a high chip length $L$ poses problems in terms of power consumption and fabrication area [13], [14]. Hence, other means to improve $\alpha$ without relying on the chip length $L$ of the PR signals are very important.

For example, suppose one is observing on-air radar signals of bandwidth up to $B = 30$ [MHz] over an extremely wide observation frequency scope $f_{max} = 40$ [GHz]. This setting is reasonable in radar systems [21], [22]. We discussed that the chip speed should not be less than the Nyquist rate, i.e., $f_c \geq f_{NYQ}$, where $f_{NYQ} = 80$ [GHz]. In this example, to achieve $f_p \approx B$, the chip length needs to be $L = 2^{11} - 1$. Although hardware implementations of such PR signal generators having $f_c = 80$ [GHz] and chip length greater than $L = 2^{11} - 1$ were proposed in the literature [23], [24], they require very large fabrication areas and high power consumption, which has hindered practical uses thus far.

### E. Problem formulation

The goal of this paper is to introduce the proposed sampling system, Section III, which aims to improve the sampling efficiency $\alpha$ with given and fixed specifications $f_p$, $f_c$, and $L$ for PR signal generation. Throughout this paper, we assume small $L$ and $B$ and a large $f_{NYQ} = f_c$, which implies $f_p$ large enough compared to $B$ and makes room for improving $\alpha$. That is, $f_p \geq pB$ for a natural number $p > 1$. Then, improving $\alpha$ can be made without upgrading the PR signal generators and causing the said implementation issues such as higher power consumption and larger fabrication area discussed in the previous subsection. Thus, very wideband signals can be losslessly sampled using commercially available PR signal generators and ADCs, while this was not possible in the past with the conventional cMWC system.

## III. ALIASED MWC

We propose *Aliased modulated wideband converters* (AMWC). AMWC renders the anti-aliasing rule $f_s \geq W_{LPF}$ used in cMWC unnecessary, as revealed later. Instead, AMWC intentionally induces and exploits aliasing at the ADC to regulate the splitting interval $f_I$ and improve $\alpha$ without relying on the specification of PR signals.

In this section, we first discuss our method to induce controlled aliasing at the ADC and derive revised input-output relationships of AMWC. We then investigate how to control the aliasing for lossless sampling. Finally, we compare the sampling efficiency of AMWC with that of cMWC.

### A. Intentional Aliasing Method

The AMWC system is depicted in Fig. 1. As mentioned already, compared to cMWC, AMWC is designed to not satisfy the anti-aliasing rule at the ADC; rather, it is designed to induce *intentional* aliasing by setting the bandwidth of LPF greater than the sampling rate. In fact, in both cMWC and AMWC, an aliasing is introduced first by the mixer. The effect of this first aliasing is shown in (2), where the mixer shifts, gives weights, and has the signal spectrum $X(f)$ overlapped with shifted versions of itself at intervals of $f_p$. By the second aliasing at the ADC, the overlapped spectrum is aliased again at intervals of new sampling rate of AMWC $f'_s$, which is smaller than the filter bandwidth. By adjusting the relationship between $f_p$ and $f'_s$, the splitting interval $f_I$, which is the interval at which $X(f)$ is split in the outputs of AMWC, is regulated.

Specifically, we set the new sampling rate $f'_s$ of AMWC:

$$f'_s = \frac{q'}{p} f_p, \tag{6}$$

where $q'$ is the new channel trading parameter for AMWC and an odd number. The bandwidth of LPF is $W_{LPF} = q'f_p$, and therefore, $W_{LPF} = pf'_s$ for the integer *aliasing parameter* $p > 1$. We will show that coprime $p$ and $q'$ with $q' > p$ is necessary for no information loss of $X(f)$. The new sampling rate induces additional aliasing and regulates the splitting interval $f_I$ to improve the sampling efficiency. We let

$$f'_p \triangleq \frac{f_p}{p} \tag{7}$$

denote *the least common shifting interval* (LCS), which will become the splitting interval of AMWC, i.e., $f_{I,AMWC} = f'_p$.

With the introduction of new sampling rate $f'_s$ in (6), it becomes easier to compare AMWC with cMWC. Specifically, with the sampling rate fixed, the number of equations for the input reconstruction obtained by cMWC and that by AMWC can be compared; with the number of equations fixed, the sampling rates for the two can be compared. For a given sampling rate $f'_s = q'f_p/p$, we will show in this section, the number of equations obtained by AMWC is $Mq'$. For a given sampling rate $f_s = qf_p$, from Section II-B, the number of equations obtained by cMWC is $Mq$. With the sampling rate fixed the same, i.e., $f_s = f'_s$, we note that $q' = qp$. This implies that AMWC has $p$-times more equations than that of cMWC. TABLE I presents an example of the increase in the number of equations of AMWC. With the number of equations fixed, i.e., $Mq = Mq'$, on the other hand, AMWC requires $p$-times smaller sampling rate than cMWC does.

In [18], a variation of MWC using a sampling rate similar to (6) was considered, to analyze the noise factor incurred by the aliasing of subbands. There appear coprime relations between $p$ and $q'$ similar to that in this paper. However, the purpose of using coprime $p$ and $q'$ in [18] was completely different from that of this paper, i.e., they regulated the splitting interval of the subbands to make a fair comparison with other filterbank-based sampling systems with regard to the effect of noise. No relation between $p$ and $q'$ for lossless sampling and improving sampling efficiency was studied in [18].

TABLE I
PARAMETER COMPARISONS BETWEEN AMWC AND cMWC

| Multiband model | | |
|---|---|---|
| $f_{NYQ} = 18$ [GHz] | $B = 30$ [MHz] | $K_B = 10$ |
| **System specification** | | |
| $L = 2^7 - 1$ | $f_p = 142$ [MHz] | $M = 3$ |
| **Parameters** | cMWC | AMWC (with $p = 4$) |
| Channel-trading parameter | $q = 5$ | $q' = 19$ |
| Sampling rate [MHz] | $f_s = f_p q = 710$ | $f'_s = f_p q'p^{-1} = 674.5$ |
| Splitting interval [MHz] | $f_I = f_p = 142$ | $f_I = f_p p^{-1} = 35.5$ |
| Sparsity | $K \leq 2K_B = 20$ | $K \leq 2K_B = 20$ |
| Number of rows of **X** | $N = L = 127$ | $N = Lp = 508$ |
| Total number of equations | $Mq = 15$ | $Mq' = 57$ |

To support intentional aliasing, AMWC requires an ADC with an operating bandwidth wider than its sampling rate. Such an ADC can be implemented by using a wideband track-and-hold amplifier (THA) developed by Hittite Corp. for the applications of EW and ELLINT in [25]. This THA has an 18 GHz bandwidth and can be integrated at the front end of commercially available ADCs of sampling rate up to 4 giga-samples per second.

To show that the AMWC obtains $Mq'$ equations, we observe the input-output relationships of the aliased samples $\tilde{y}_i[n]$ in Fig. 1. Without loss of generality, we assume $q' = q$ and $f_s = pf'_s$. By the sampling theorem, the DTFT of $\tilde{y}_i[n]$ is the sum of shifts of $Y_i(f)$:

$$\tilde{Y}_i\left(e^{j2\pi fT'_s}\right) = \sum_{r=-\infty}^{\infty} Y_i\left(f - rf'_s\right)$$
$$= \sum_{r=-\infty}^{\infty} \sum_{l=-\infty}^{\infty} c_{i,l} X\left(f - rf'_s - lf_p\right) H\left(f - rf'_s\right), \tag{8}$$

where $T'_s \triangleq (f'_s)^{-1}$ and $Y_i(f)$ given in (2) is the spectrum of the output of the LPF $H(f)$. Within only a single period of $\tilde{Y}_i\left(e^{j2\pi fT'_s}\right)$ in (8), i.e., $\mathcal{F}_s(f_0) \triangleq [f_0, f_0 + f'_s)$ for any $f_0 \in \mathbb{R}$, because the bandwidth of $Y_i(f)$ is limited by the LPF $H(f)$, most of the shifts $Y_i\left(f - rf'_s\right)$ for sufficiently large $|r|$ are zeros. In other words, there exist $(f_0, R_1, R_2)$ such that the infinite order of the outer summation in (8) is reduced to a finite order, i.e.,

$$\tilde{Y}_i\left(e^{j2\pi fT'_s}\right) = \sum_{r=R_1}^{R_2} \sum_{l=-\infty}^{\infty} c_{i,l} X\left(f - rf'_s - lf_p\right) H\left(f - rf'_s\right) \tag{9}$$

for $f \in \mathcal{F}'_s(f_0)$. Assuming $H(f) = 1$ for $f \in \mathcal{F}_{LPF}$, if $f_0$, $R_1$, and $R_2$ satisfy the conditions of Lemma 1, the LPF responses in (9) are replaced with $H\left(f - rf'_s\right) = 1$ for $f \in \mathcal{F}_s(f_0)$. Note that, when $p = 1$, i.e., no aliasing exists at the ADC, $R_1 = R_2$, which is equivalent to cMWC.

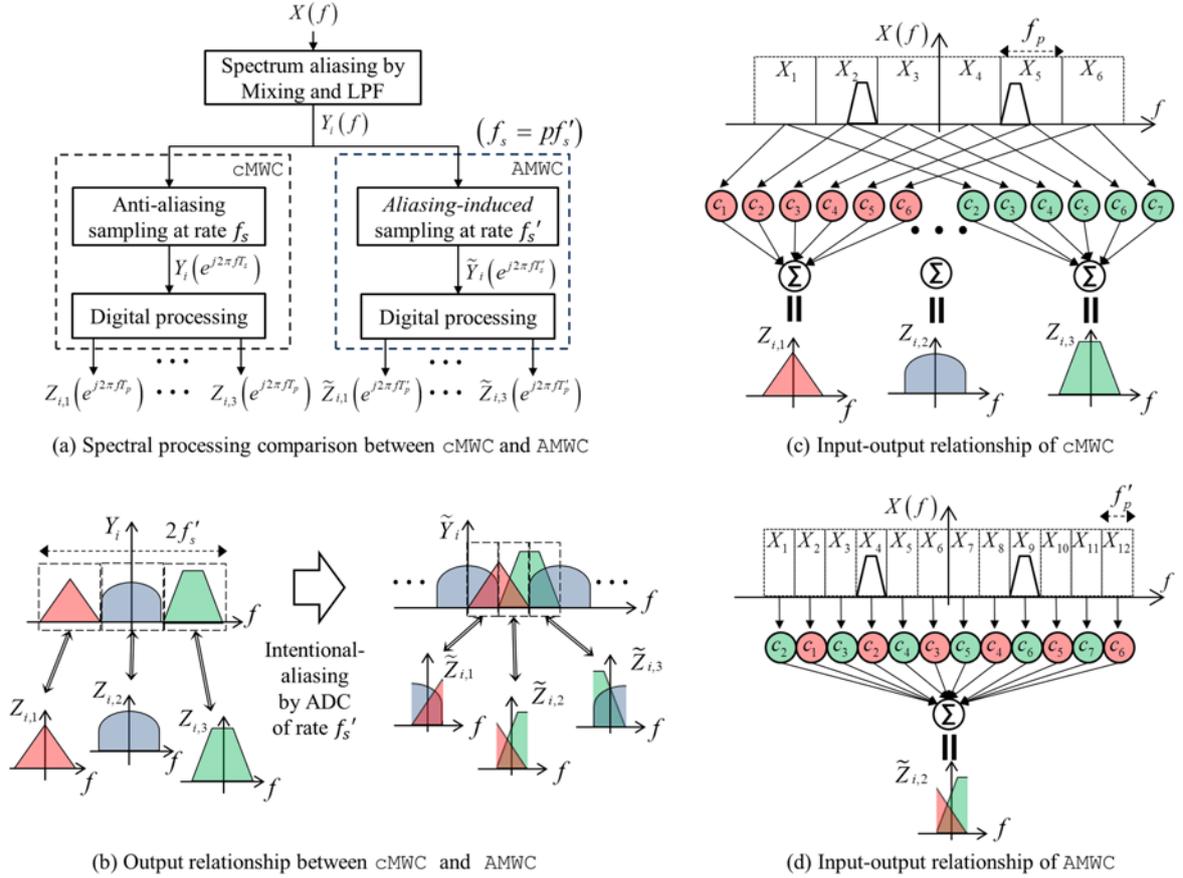

(a) Spectral processing comparison between cMWC and AMWC

(b) Output relationship between cMWC and AMWC

(c) Input-output relationship of cMWC

(d) Input-output relationship of AMWC

Fig. 3. Principle of improving the sampling efficiency by AMWC at a single analog channel is illustrated, with setting $q=3$, $q'=q$, $p=2$, and $m=3$. At the first stage, the input spectrum $X(f)$ is aliased by mixing it with the PR signal and low-pass filtering it. This aliased-version of $X(f)$ is depicted as $Y_i(f)$. In (a), the main difference between cMWC and AMWC is how to take time-samples of $Y_i(f)$. cMWC prevents the spectrum from being aliased in taking time-samples. AMWC, on the contrary, aims to make the spectrum $Y_i(f)$ intentionally aliased once again, as depicted as $\tilde{Y}_i(f)$ in (b). In (c), as a result, the splitting-interval of cMWC is $f_p$, whereas in (d), that of AMWC is halved to $f_p'$. Thus, the sampling efficiency of AMWC becomes doubled (as $p=2$).

**Lemma 1.** Equation (9) is equivalent to (8) if $f_0$, $R_1$, and $R_2$ with $R_1 < R_2 \in \mathbb{Z}$ satisfy

$$R_2 - R_1 = p - 1, \tag{10}$$

and

$$f_0 = \left(R_2 - \frac{p}{2}\right)f_s'. \tag{11}$$

*Proof:* See Appendix A.

We represent the shifting indices $rf_s' + lf_p$ in (9) in terms of the LCS $f_p'$. Then,

$$\tilde{Y}_i\left(e^{j2\pi f f_s'}\right) = \sum_{r=R_1}^{R_1+p-1}\sum_{l=-\infty}^{\infty} c_{i,l}X\left(f-(rq'+lp)f_p'\right) \tag{12}$$

for $f \in \mathcal{F}_s'(f_0)$. To merge the inner and outer summations in (12), we use Lemma 2.

**Lemma 2.** If $p$ and $q'$ are coprime, the linear combination $rq' + lp$ for $r \in \mathcal{P} \triangleq \{R_1, \cdots, R_1 + p - 1\}$ and $l \in \mathbb{Z}$ spans every integer.

*Proof:* We consider the following congruent relationship

$$k \equiv rq' \,(\mathrm{mod}\, p). \tag{13}$$

By modular arithmetic, if $p$ and $q'$ are coprime, there always exists one-to-one correspondence between $r$ and $k$ in the least residue system modulo $p$. Since $|\mathcal{P}| = p$, $rq'\,(\mathrm{mod}\, p)$ for $r \in \mathcal{P}$ in (13) spans every number in the least residue system of modulo $p$. Hence, for $r \in \mathcal{P}$ and $l \in \mathbb{Z}$, $rq' + lp = k \,\mathrm{mod}\, p + lp$ spans every integer. ∎

By denoting $k = rq' + lp$ in (12), we have the equivalent relationship

$$\tilde{Y}_i\left(e^{j2\pi f f_s'}\right) = \sum_{k=-\infty}^{\infty} d_{i,k}\left(R_1, p, q'\right)X\left(f - kf_p'\right) \tag{14}$$

for $f \in \mathcal{F}_s'(f_0)$, where $d_{i,k}(R_1, p, q')$ are the new sensing coefficients of AMWC. Proposition 3 provides the rule to obtain the coefficients $d_{i,k}$ from the Fourier coefficients $c_{i,l}$ of PR signals.

**Proposition 3.** For coprime $p$ and $q'$, let us define

$$I(k;R_1,p,q')\triangleq\frac{1}{p}\left\{k-q'\left[\left((q')^{-1}k-R_1\right)\bmod p+R_1\right]\right\}, \quad (15)$$

where $(q')^{-1}(\bmod\,p)$ is the multiplicative inverse of $q'$ modulo $p$. Equation (14) is equivalent to (12) if

$$d_{i,k}(R_1,p,q')=c_{i,I(k;R_1,p,q')}. \quad (16)$$

*Proof:* See Appendix B.

In (14), the bandwidth of the subbands $X\left(f-kf_p'\right)$ for $f\in\mathcal{F}_s'(f_0)$ equals $f_s'$ and is $q'$ times wider than their shifting interval $f_p'$. Therefore, every subband is correlated with the closest $q'-1$ subbands. By making these subbands spectrally orthogonal, the $M$ relationships for $i=1,\cdots,M$ are expanded to $Mq'$ equations to enhance the input reconstruction performance. A similar work was done for cMWC through (3) to (4), which further divides the observing frequency domain $\mathcal{F}_s'(f_0)$ (14) into $q'$ tiny domains. Specifically, for $u=0,\cdots,q'-1$, the $u$-th tiny frequency domain is defined by $\mathcal{F}_p'\left(f_0+uf_p'\right)$, where

$$\mathcal{F}_p'(f_0)\triangleq\left[f_0,f_0+f_p'\right). \quad (17)$$

Then, the corresponding divided outputs have relationships

$$\begin{aligned}
&\widetilde{Y}_i^{(u)}\left(e^{j2\pi f f_p'}\right)\\
&=\sum_{k=-\infty}^{\infty}d_{i,k}(R_1,p,q')X\left(f-kf_p'\right)\text{ for }f\in\mathcal{F}_p'\left(f_0+uf_p'\right),
\end{aligned} \quad (18)$$

for $u=0,\cdots,q'-1$. Finally, we define the output $\widetilde{Z}_{i,u}\left(e^{j2\pi f f_p'}\right)$ of AMWC as follows:

$$\begin{aligned}
\widetilde{Z}_{i,u}\left(e^{j2\pi f f_p'}\right)&\triangleq\widetilde{Y}_i^{(u)}\left(e^{j2\pi f f_p'}\right)\Big|_{f=f+uf_p'}\\
&=\sum_{k=-\infty}^{\infty}d_{i,k+u}(R_1,p,q')X\left(f-kf_p'\right)
\end{aligned} \quad (19)$$

for $f\in\mathcal{F}_p'(f_0)$. The final output $\widetilde{z}_{i,u}\left[\widetilde{n}\right]$ in the discrete-time domain can be obtained by performing digital frequency modulation and low-pass filtering on $\widetilde{y}_i[n]$, as similarly done for cMWC in (3). The specific design of the digital processing system is shown in Fig. 1-(b).

Consequently, in (19), the input $X(f)$ is split into spectrally orthogonal subbands at intervals of $f_p'$. Therefore, the splitting interval of AMWC equals the LCS $f_p'$:

$$f_{I,AMWC}=f_p'\triangleq\frac{f_p}{p}, \quad (20)$$

which is $p$ times lower than $f_{I,cMWC}$. By reducing the splitting interval by controlling the aliasing parameter $p$, the sampling efficiency of AMWC in (5) is improved. Fig. 3 illustrates how AMWC regulates the splitting interval and improves the sampling efficiency. In contrast, as discussed in Section II-D, regulating the splitting interval of cMWC requires a very costly solution of advanced PR signal generators with a larger chip length. Consequently, both cMWC and AMWC obtain $Mq=Mq'$ equations for input reconstruction, although AMWC consumes a $p$-times lower total sampling rate (6). In Section III-C, we show that the $Mq'$ equations of AMWC are independent.

### B. Matrix Form of Input–Output Relationship

For convenience of analyzing and solving linear simultaneous $Mq'$ equations (19), we cast them as a matrix equation. To this end, we first reduce the infinite summation in (19) to be finite. We then discretize the continuous spectra to form a matrix with a finite number of columns.

Since $X(f)$ is band-limited to $f\in\mathcal{F}_{NYQ}$, within the limited frequency range $f\in\mathcal{F}_p'(f_0)$, the infinite summation order in (19) is reduced to a finite order as follow:

$$\begin{aligned}
&\widetilde{Z}_{i,u}\left(e^{j2\pi f f_p'}\right)\\
&=\sum_{k=N_1}^{N_2}d_{i,k+u}(R_1,p,q')X\left(f-kf_{I,AMWC}\right)\text{ for }f\in\mathcal{F}_p'(f_0),
\end{aligned} \quad (21)$$

where $N_1$ and $N_2$ are, respectively, the smallest and largest index $k$ of the subbands $X\left(f-kf_{I,AMWC}\right)$ that contain some active value of $X(f)$ within $f\in\mathcal{F}_{NYQ}$. Namely, these indices $N_1$ and $N_2$ indicate $X\left(f-kf_{I,AMWC}\right)=0$ for $k<N_1$ and $k>N_2$, and thus help us obtain a matrix equation of (21) with finite dimensions. To mathematically define $N_1$ and $N_2$, note that the $k$-th subband $X\left(f-kf_{1,AMWC}\right)$ in (21) observes the frequency range

$$\mathcal{F}_k \triangleq \left[f_0 - kf_{1,AMWC}, f_0 - kf_{1,AMWC} + f'_p\right] \quad (22)$$

of $X(f)$. Then, the indices $N_1$ and $N_2$ are defined by

$$\begin{aligned} N_1 &\triangleq \min\left\{k \in \mathbb{Z} : \mathcal{F}_k \cap \mathcal{F}_{NYQ} \neq \varnothing\right\} \\ &= \min\left\{k \in \mathbb{Z} : f_0 - kf_{1,AMWC} < f_{\max}\right\} \end{aligned} \quad (23)$$

and

$$\begin{aligned} N_2 &\triangleq \max\left\{k \in \mathbb{Z} : \mathcal{F}_k \cap \mathcal{F}_{NYQ} \neq \varnothing\right\} \\ &= \max\left\{k \in \mathbb{Z} : f_0 - kf_{1,AMWC} + f'_p > -f_{\max}\right\}, \end{aligned} \quad (24)$$

respectively. Using the parameters and relations given in TABLE II and Lemma 1, the two problems (23) and (24) turn into

$$N_1 = \min\left\{k \in \mathbb{Z} : R_2 q' - \frac{(q'+L)p}{2} < k\right\} \quad (25)$$

and

$$N_2 = \max\left\{k \in \mathbb{Z} : R_2 q' - \frac{(q'-L)p}{2} + 1 > k\right\} \quad (26)$$

respectively. As both $q'$ and $L$ are odd positive integers, the solutions of two problems (25) and (26) are determined as follow:

$$N_1 = R_2 q' - \frac{(q'+L)p}{2} + 1, \quad (27)$$

and

$$N_2 = R_2 q' - \frac{(q'-L)p}{2}. \quad (28)$$

Finally, the output spectrum $\widetilde{Z}_{i,u}\left(e^{j2\pi f T'_p}\right)$ in (21) turns into a linear combination of unknown subbands $X\left(f-kf_{1,AMWC}\right)$ for $f \in \mathcal{F}'_p(f_0)$. The matrix-multiplication form $\mathbf{Z} = \mathbf{DX}$ of (21) is provided in (30). We denote the number of subbands, i.e., the dimension of matrix $\mathbf{X}$, by $N$, which equals

$$\begin{aligned} N &= N_2 - N_1 + 1 \\ &= Lp. \end{aligned} \quad (29)$$

Since $X(f)$ consists of $K_B$ narrow bands over the wide

Nyquist range, only a few of its subbands $X\left(f-kf_{1,AMWC}\right)$ for $f \in \mathcal{F}'_p(f_0)$ have nonzero values. Therefore, the matrix $\mathbf{X}$ in (30) is row-wise sparse with a sparsity $K$ related to $K_B$.

To draw a relationship between the analytic result (30) and actually acquired samples $\widetilde{z}_{i,u}[\widetilde{n}]$, we convert the DTFT (30) to the DFT of $\widetilde{z}_{i,u}[\widetilde{n}]$ by taking the frequency samples of the infinite columns of $\mathbf{Z}$ and $\mathbf{X}$. When the input is observed for a finite duration $T_o$, taking samples of the spectrum (21) at frequency intervals of $\Delta f = T_o^{-1}$ does not cause any information loss. The samples of spectrum $\widetilde{Z}_{i,u}\left(e^{j2\pi f T'_p}\right)$ is obtained by taking the DFT of the actually acquired time-samples $z_{i,u}[\widetilde{n}]$. Consequently, for a finite observation time $T_o = 2WT'_p$ for a sample length $2W$, we rewrite the matrix-multiplication form (30) as

$$\mathbf{Z}_{2W} = \mathbf{DX}_{2W}, \quad (31)$$

where columns of $\mathbf{Z}_{2W} \in \mathbb{C}^{Mq \times 2W}$ and $\mathbf{X}_{2W} \in \mathbb{C}^{N \times 2W}$ are sub-columns of $\mathbf{Z}$ and $\mathbf{X}$, respectively, at frequency intervals of $\Delta f$. This concept will be exploited in Section IV to derive a revised input-output relationship of AMWC for using LPF with a non-flat frequency response.

### C. Choosing the Aliasing Parameter

For a given total sampling rate, AMWC obtains more equations used for input reconstruction than cMWC does. What remains is to check if the extended equations provide independent information. We reveal a condition on the aliasing parameter $p$ that necessitates the linear system (30) to be well-posed for every $K$-sparse signal matrix $\mathbf{X}$.

**Proposition 4.** There exists the unique solution of (30) for every $K$-sparse signal $\mathbf{X}$ only if $p$ and $q'$ are coprime and $q' > p$.

*Proof:* See Appendix C.

Proposition 4 gives a condition $q' < p$ for coprime $p$ and $q'$ that makes AMWC an ill-posed system. This indicates that, within the set of coprime $q' > p$, there may be a subset that makes AMWC guarantees the existence of unique solution of (30) for every $K$-sparse signal matrix $\mathbf{X}$.

In [11], a CS result states there exist the unique solution of a multiple measurement vector (MMV) CS equation $\mathbf{Z} = \mathbf{DX}$ for every $K$-sparse signal $\mathbf{X}$ if

$$
\underbrace{\begin{pmatrix} \widetilde{Z}_{1,0}\left(e^{j2\pi f T'_p}\right) \\ \vdots \\ \widetilde{Z}_{1,q'-1}\left(e^{j2\pi f T'_p}\right) \\ \widetilde{Z}_{2,0}\left(e^{j2\pi f T'_p}\right) \\ \vdots \\ \widetilde{Z}_{M,q'-1}\left(e^{j2\pi f T'_p}\right) \end{pmatrix}}_{\triangleq \mathbf{Z} \in \mathbb{C}^{Mq \times \infty}}
=
\underbrace{\begin{pmatrix} d_{1,N_1} & d_{1,N_1+1} & \cdots & d_{1,N_2} \\ \vdots & \vdots & & \vdots \\ d_{1,N_1+(q'-1)} & d_{1,N_1+(q'-1)+1} & \cdots & d_{1,N_2+(q'-1)} \\ d_{2,N_1} & d_{2,N_1+1} & \cdots & d_{2,N_2} \\ \vdots & \vdots & & \vdots \\ d_{M,N_1+(q'-1)} & d_{M,N_1+(q'-1)+1} & \cdots & d_{M,N_2+(q'-1)} \end{pmatrix}}_{\triangleq \mathbf{D} \in \mathbb{C}^{Mq \times N}}
\underbrace{\begin{pmatrix} X\left(f - N_1 f'_p\right) \\ X\left(f - (N_1+1)f'_p\right) \\ \vdots \\ X\left(f - N_2 f'_p\right) \end{pmatrix}}_{\triangleq \mathbf{X} \in \mathbb{C}^{N \times \infty}}
\quad (30)
$$

$$2K < \text{spark}(\mathbf{D}) - 1 + \text{rank}(\mathbf{X}), \qquad (32)$$

where *spark* is the minimum number of linearly dependent columns in $\mathbf{D}$. Meanwhile, the spark of an $Mq'$-by-$N$ matrix is upper bounded to $Mq'+1$ by the Singleton bound [26]. Based on these results, we find a sufficient condition on $p$ and $q'$ from Monte Carlo experiments in Section V-A (Fig. 4) that maximizes the spark of $\mathbf{D}$.

**Main Result 5.** Let $Mq' \geq 2K$. For every $K$-sparse signal $\mathbf{X}$, there exists the unique solution of (30), and therefore, AMWC does not lose any information of $K$-sparse signal $\mathbf{X}$, if $p$ and $q'$ are coprime and $q' > p$.

Meanwhile, we choose $p$ to minimize the maximum of the sparsity $K$, which is the number of nonzero subbands of $X(f)$ at splitting intervals $f_{I,AMWC} = f_p'$. The sparsity $K$ is dependent on the center frequencies of $K_B$ multibands and their maximum bandwidth $B$. When $f_{I,AMWC} \geq B$, every multiband occupies at most two subbands, which implies $K \leq 2K_B$. On the other hand, when $f_{I,AMWC} < B$, some multibands may occupy more than two subbands, which provides an opportunity to increase $K$ beyond $2K_B$. Hence, we recommend choosing the aliasing parameter $p$ as

$$p \leq \left\lfloor \frac{f_p}{B} \right\rfloor. \qquad (33)$$

### D. Sampling Efficiency Analysis

We compare the sampling efficiencies of AMWC, $\alpha_{AMWC}$, and cMWC, $\alpha_{cMWC}$, defined in (5). The sampling efficiencies are functions of the sparsity $K$, which is a random variable in general. We denote the sparsity of cMWC and AMWC by $K_{cMWC}$ and $K_{AMWC}$, respectively. To make them deterministic, we put assumptions on $K_{cMWC}$ and $K_{AMWC}$ that in both cMWC and AMWC, the $K_B$ bands in $X(f)$ respectively occupies exactly one subband, i.e., $K_{cMWC} = K_{AMWC} = K_B$. This occurs with high probability when $f_p p^{-1} \gg B$ and the center frequencies of multibands are far enough apart from each other with a small $K_B$.

Under the assumption above, the sampling efficiencies of cMWC and AMWC are obtained by

$$\alpha_{cMWC} = \frac{K_B B}{K_{cMWC} f_{I,cMWC}} = \frac{B}{f_p}, \qquad (34)$$

and

$$\alpha_{AMWC} = \frac{K_B B}{K_{AMWC} f_{I,AMWC}} = \frac{pB}{f_p}, \qquad (35)$$

respectively. Note that if $p = 1$, AMWC and cMWC are completely identical, and therefore $\alpha_{AMWC} = \alpha_{cMWC}$. When $p > 1$, the intentional aliasing of AMWC takes effect and improves the sampling efficiency proportionally to $p$.

## IV. NON-IDEAL LOW-PASS FILTERS

The input-output relationship in the previous section is based on the ideal LPF $H(f)$ having a flat pass-band response. However, in real applications, the pass-band response of an LPF significantly fluctuates. In the case of cMWC, a post digital-processing technique to equalize the effects of non-flat filter responses was proposed in [27]. Unfortunately, owing to the aliasing at ADC, the equalizations cannot be applied to AMWC. In this section, we instead provide a revised input-output relationship of AMWC based on the fluctuated LPF $G(f)$. Without loss of generality, we assume all analog channels use the same LPF. We assume that the response $G(f)$ is nonzero and known within the pass-band $f \in \mathcal{F}_{LPF}$ and is zero for $f \in \mathcal{F}_{LPF}{}^C$. We derive a revised input-output relationship reflecting the effect of $G(f)$. Paradoxically, our empirical results in Section V conclude that, for a given sampling efficiency, an irregularly fluctuated filter response is helpful to further decrease the total sampling rate required for lossless sub-Nyquist sampling.

The derivation starts from substituting $H(f)$ in the input-output relations of (8)-(12) with $G(f)$. Without loss of generality, we assume $q' = q$ and $f_s = pf_s'$. Equation (9) then turns into

$$\widetilde{Y}_i\left(e^{j2\pi f f_s'}\right) = \sum_{r=R_1}^{R_2} \sum_{l=-\infty}^{\infty} c_{i,l} X\left(f - (rq'+lp)f_p'\right) G\left(f - rq'f_p'\right) \qquad (36)$$

for $f \in \mathcal{F}_s'(f_0)$, where $R_1$ and $R_2$ are chosen from Lemma 1. By Lemma 2, we substitute $rq' + lp = k$ and merge the outer and inner summations:

$$\widetilde{Y}_i\left(e^{j2\pi f f_s'}\right) = \sum_{k=N_1}^{N_2} d_{i,k}(R_1, p, q') X\left(f - kf_p'\right) G\left(f - \gamma_p(k)f_p'\right) \qquad (37)$$

for $f \in \mathcal{F}_s'(f_0)$, where the sensing coefficients $d_{i,k}(R_1, p, q')$, $N_1$, and $N_2$ are, respectively, computed from Proposition 3, (27), and (28). We define the function $\gamma_p$ of $k$ that maps $k$ in (37) to the corresponding $rq'$ in (36) so that the two equations are equivalent. Lemma 6 reveals the mapping rule for $\gamma_p(k)$.

**Lemma 6.** Under the conditions of Lemma 1 and Lemma 2, (36) and (37) are equivalent if the mapping rule of $\gamma_p$ is assigned by

$$\gamma_p(k) = k - pI(k; R_1, p, q'), \qquad (38)$$

where the picking regularity $I(k; R_1, p, q')$ is defined in (15).

*Proof:* See Appendix B.

As done in (14) to (19), the final outputs $\tilde{z}_{i,u}\left[\tilde{n}\right]$ for $u = 0, \cdots, q'-1$ are obtained by processing the time-samples $\tilde{y}_i[n]$ of the spectrum (37) using the digital system given in Fig. 1-(b). Then, those spectra $\widetilde{Z}_{i,u}\left(e^{j2\pi fT'_p}\right)$ have the following input-output relationships:

$$
\begin{aligned}
&\widetilde{Z}_{i,u}\left(e^{j2\pi fT'_p}\right) \\
&= \sum_{k=N_1}^{N_2} d_{i,k+u}\left(R_1,p,q'\right)X\left(f-kf'_p\right)G\left(f+uf'_p-\gamma_p\left(k+u\right)f'_p\right) \\
&= \sum_{k=N_1}^{N_2} d_{i,k+u}\left(R_1,p,q'\right)G\left(f-\gamma'_p\left(k,u\right)f'_p\right)X\left(f-kf'_p\right)
\end{aligned}
\tag{39}
$$

for $f\in\mathcal{F}'_p\left(f_0\right)$, where $\gamma'_p\left(k,u\right)\triangleq\gamma_p\left(k+u\right)-u$.

Consequently, the linear coefficients on the subbands $X\left(f-kf'_p\right)$ in (39) become frequency-selective. To numerically solve (39), we discretize the continuous frequency, as discussed in Section III-B. We assume that the signal is observed for the finite duration $T_o=2WT'_p$, where $2W$ is the length of the discretized signal. Then, the samples of spectrum are defined by

$$
\begin{aligned}
\widetilde{Z}_{i,u}\left[w\right] &\triangleq \widetilde{Z}_{i,u}\left(e^{j2\pi fT'_p}\right)\Big|_{f=wT_o^{-1}} \\
&= \sum_{k=N_1}^{N_2} d_{i,k+u}\left[G\left(f-\gamma'\left(k,u\right)f'_p\right)X\left(f-kf'_p\right)\right]_{f=wT_o^{-1}} \\
&= \sum_{k=N_1}^{N_2} b_{(i,u),k}\left[w\right]X\left(f-kf'_p\right)_{f=wT_o^{-1}}
\end{aligned}
\tag{40}
$$

for $w\in\mathcal{W}\triangleq\left\{f_0T_o,\cdots,\left(f_0+f'_p\right)T_o-1\right\}$, where the frequency-selective sensing coefficients $b_{(i,u),k}\left[w\right]$ are defined as

$$
b_{(i,u),k}\left[w\right]\triangleq d_{i,k+u}G\left(f-\gamma'\left(k,u\right)f'_p\right)_{f=wT_o^{-1}}
\tag{41}
$$

for $w\in\mathcal{W}$. Note that, by the relation between DFT and DTFT, the spectrum samples (40) are obtained by taking the DFT as follows:

$$
\widetilde{Z}_{i,u}\left[w\right]=\sum_{n=0}^{2W-1}\widetilde{z}_{i,u}\left[\widetilde{n}\right]e^{j2\pi\frac{\widetilde{n}}{2W}\left(w\bmod 2W\right)} \text{ for } w\in\mathcal{W},
\tag{42}
$$

where $\widetilde{z}_{i,u}\left[\widetilde{n}\right]$ are the output sequences of AMWC.

For convenience, we represent the input-output relation of (40) for $w\in\mathcal{W}$ in a vector form as

$$
\widetilde{\mathbf{Z}}\left[w\right]=\mathbf{B}\left[w\right]\mathbf{X}\left[w\right],
\tag{43}
$$

where the elements of the output column vector $\widetilde{\mathbf{Z}}\left[w\right]\in\mathbb{C}^{Mq'}$ are $\widetilde{Z}_{i,u}\left[w\right]$ for row indices $i=1,\cdots,M$ and $u=0,\cdots,q'-1$. The unknown column vector $\mathbf{X}\left[w\right]\in\mathbb{C}^N$ consists of $X\left(f-kf'_p\right)_{f=wT_o^{-1}}$ for row indices $k=N_1,\cdots,N_2$. The frequency-selective sensing matrix $\mathbf{B}\left[w\right]\in\mathbb{C}^{Mq'\times N}$ consists of $b_{(i,u),k}\left[w\right]$ with row indices $i$ and $u$ and column index $k$.

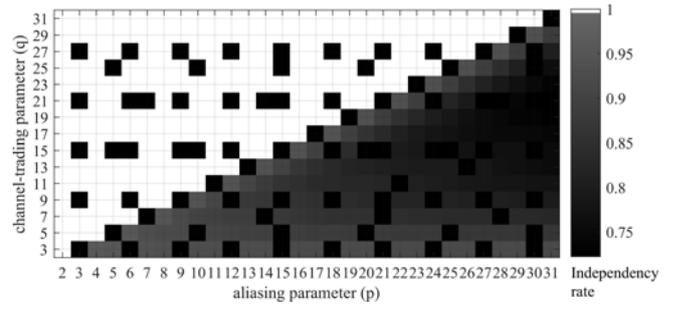

Fig. 4. Independency rates under various $p$ and $q'$ for which randomly selected $Mq'$ columns of the sensing matrix $\mathbf{D}\in\mathbb{C}^{Mq'\times N}$ of AMWC are independent. When $p$ and $q'$ are coprime and $q'>p$, every selection of $Mq'$ columns is linearly independent.

The CS model (43) is called MMV with different sensing matrices, for which many numerical solvers have been developed [9], [28].

The existence of unique solution of (43) depends on the spark of sensing matrix $\mathbf{B}\left[w\right]$. Note that from (41), the elements of $\mathbf{B}\left[w\right]$ are multiplications of the elements of $\mathbf{D}$ and the samples of the low pass filter $G\left(f\right)$. In [29], Davies *et al.* proved that the spark of a matrix from an independent continuous distribution achieves the Singleton bound with probability one. When the filter response $G\left(f\right)$ is designed to be irregular, i.e., its samples are drawn from an independent random distribution, the spark of $\mathbf{B}\left[w\right]$ after multiplication with the samples of $G\left(f\right)$ should grow closer to achieving the Singleton bound. When the spark of $\mathbf{B}\left[w\right]$ indeed achieves the Singleton bound and the condition (32) holds, for every $K$-sparse signal $\mathbf{X}$ the unique solution to (43) always exists.

## V. SIMULATION

### A. Spark of Sensing Matrix

To support Main Result 5, the sufficiency of lossless sub-Nyquist sampling by AMWC, we demonstrate that the sensing matrix $\mathbf{D}$ with coprime parameters $q'>p$ achieves the Singleton bound.

Monte Carlo experiments were performed under various settings of $p$ and $q'$. With $L=127$, we used the maximum length sequences of length $L$ as the chip values of PR signal for each channel $i=1,\cdots,M$. We set the number of analog channels to $M=3$. For $5\times10^5$ independent trials, we randomly selected $Mq'$ columns of $\mathbf{D}$ and counted the rate for which the selected columns are linearly independent.

Fig. 4 shows how the linear independency of columns in $\mathbf{D}$ varies as $p$ and $q'$ change. The white points in the plot indicate the pairs of $p$ and $q'$ where every selection of $Mq'$ columns of $\mathbf{D}$ is linearly independent. The dark points indicate that at least one selection of $Mq'$ columns has linear dependency. The upper triangular area indicates the region of $\left(p,q'\right)$ with $q'>p$ where all points except for the points that $p$ and $q'$ are not coprime belong to the white set. That is, for coprime $q'>p$, all the selections of $Mq'$ columns are linearly independent, and thus the spark of $\mathbf{D}$ achieves the

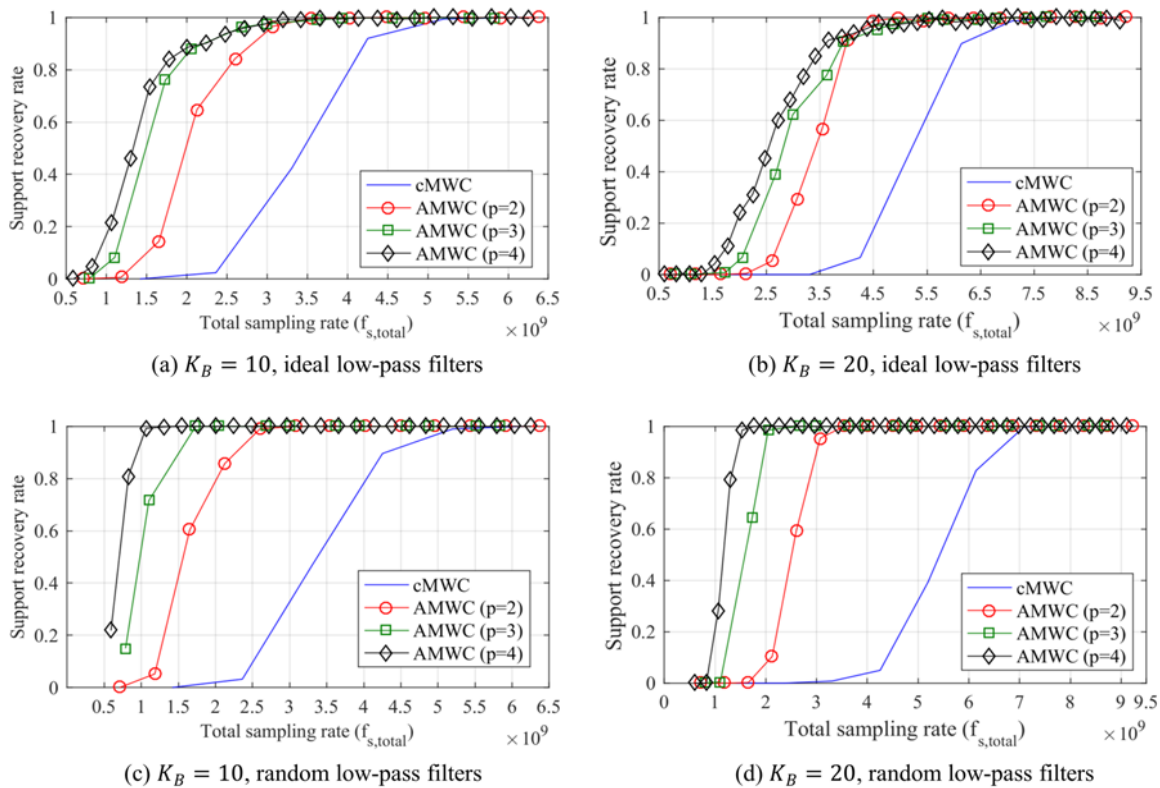

Fig. 5. Rate of successful support recovery of cMWC and AMWC as a function of total sampling rate for various aliasing parameters $p$ and multibands $K_B$. The number of channels was fixed to $M = 3$. Ideal ((a)-(b)) and random ((c)-(d)) low-pass filters were used.

Singleton bound. This result is consistent with Proposition 4 and supports Main Result 5.

### B. Reduction of Total Sampling Rate

We demonstrate that, with the improved sampling efficiency, AMWC indeed reduces the total sampling rate required for lossless sub-Nyquist sampling for given specifications of PR signals. Additionally, when the frequency response of low-pass filters is drawn at random, the reduction of total sampling rate is boosted. The reduction of total sampling rate reduces the number of channels as well as the sampling rate of each channel.

For simulation, we generated real-valued multiband inputs $x(t)$ as the sum of $K_B$ narrow band signals of bandwidth $B = 5$ [MHz]. The energies of narrow bands are equal. The center frequencies of narrow band signals were drawn at random, while those spectra were not overlapped with each other. The maximum frequency of $x(t)$ does not exceed $f_{max} = 10$ [GHz]. The signals last for the duration $T_o = 2WT_p'$ seconds with $W = 15$. The parameters of PR signals were $L = 127$, $f_p = 2f_{max}L^{-1} = 157.48$ [MHz]. We used maximum length sequences with different initial seeds as the chip values of PR signals for channel indices $i = 1, \cdots, M$. We expressed the continuous signals in simulation on a dense discrete-time grid with intervals of $\left(2q'f_{NYQ}\right)^{-1}$ seconds. The bandwidth of low-pass filters and the sampling rate followed the parameter relations of AMWC, i.e., $W_{LPF} = q'f_p$ and $f_s' = p^{-1}W_{LPF}$. We considered the ideal LPF $H(f)$ with a flat passband response and the non-ideal

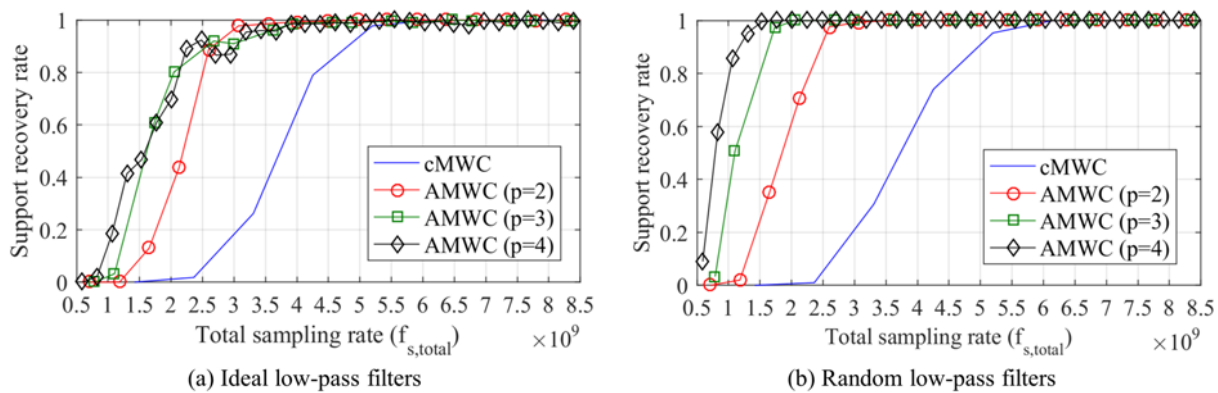

Fig. 6. Rate of successful support recovery of cMWC and AMWC as a function of total sampling rate when SNR=3 [dB]. The number of channels was fixed to $M = 3$, and the number of multibands in $X(f)$ is fixed to $K_B = 10$. Ideal (a) and random (b) low-pass filters were used.

LPF $G(f)$ with an irregular passband response. In simulation, the impulse response of $G(f)$ was drawn initially from the normal distribution, windowed to limit the filter bandwidth, and then held fixed throughout the whole simulation. We call $G(f)$ the random LPF with this irregular passband response. Under various settings of $p$, $q'$, and $K_B$ with coprime $q' > p$, we measured the rate of successful recovery of the supports of $\mathbf{X}$ by the distributed CS orthogonal matching pursuit (DCS-SOMP) algorithm [28]. For single supports estimation, DCS-SOMP was run for $2K_B$ iterations. It aimed to find one distinct support per each iteration out of $K$ supports, given $K \leq 2K_B$. Once the supports are found, $x$ can be reconstructed by the least squares. The successful support recovery was declared if $\mathcal{S} \subseteq \hat{\mathcal{S}}$, where $\mathcal{S}$ and $\hat{\mathcal{S}}$ are, respectively, the true and found supports. The support recovery rate in simulations was defined as the number of successful support recovery divided by total 500 trials with randomly regenerated $x(t)$.

Fig. 5 shows the support recovery rate of AMWC as a function of total sampling rate when $M = 3$. We set $K_B = \{10, 20\}$. Plots (a) and (b) are results of using the ideal LPF $H(f)$. It is demonstrate that compared to cMWC, AMWC reduces the total sampling rate required for reconstruction of given multiband signals. Inversely, for a given sampling rate, AMWC takes sub-Nyquist samples of more multibands than cMWC does, without information loss.

However, when $p$ increases, although the sampling efficiency is improved proportionally to $p$ from (35), the total sampling rate does not decrease anymore. This is caused by the lack of degrees of freedom in the sensing matrix $\mathbf{D}$. The elements of $\mathbf{D}$ are made of the Fourier coefficients $c_{i,j}$ of the PR signals, and most elements are repeatedly reused. Although it was demonstrated in the previous sub-section that $\mathbf{D}$ has the maximum spark and well preserves the sparse signal $\mathbf{X}$, recovering $\mathbf{X}$ by non-optimal CS algorithms requires $\mathbf{D}$ to have a large degrees of freedom [10]. This limitation is overcome by using the random LPF $G(f)$.

Plots (c) and (d) are the results of using the random LPF $G(f)$. It is shown that AMWC further reduces the total sampling rate required for successful support recovery as the sampling efficiency improves. Consequently, the random

TABLE III
THE TOTAL SAMPLING RATE REQUIRED FOR 90% SUPPORT RECOVERY RATE
WITH VARIOUS SNR AND VALUES OF $p$

| SNR [dB] | LPF | $p=1$ (cMWC) | $p=2$ (AMWC) | $p=3$ (AMWC) | $p=4$ (AMWC) |
|---|---|---|---|---|---|
| -6 | Ideal | 6.142 | 4.016 | 3.622 | 3.898 |
| | **Random** | **6.142** | **3.543** | **2.677** | **2.244** |
| -3 | Ideal | 6.142 | 3.543 | 3.622 | 3.425 |
| | **Random** | **6.142** | **3.071** | **2.047** | **1.535** |
| 0 | Ideal | 5.197 | 3.071 | 2.677 | 2.953 |
| | **Random** | **5.197** | **2.598** | **1.732** | **1.535** |
| 3 | Ideal | 5.197 | 3.071 | 2.677 | 2.480 |
| | **Random** | **5.197** | **2.598** | **1.732** | **1.299** |
| 12 | Ideal | 5.197 | 3.071 | 2.677 | 2.244 |
| | **Random** | **5.197** | **2.126** | **1.732** | **1.063** |

The floating numbers in cells indicate the minimal total sampling rate in GHz which achieves the support rate recovery of 90%. The number of analog channels and multibands were set to $M = 3$ and $K_B = 10$, respectively.

response of $G(f)$ enhances the degrees of freedom of sensing matrices $\mathbf{B}[w]$ for different frequency indices $w$ and improves the recovery performance by the non-optimal algorithm DCS-SOMP. This enhancement cannot be applied for cMWC, since the effect of random response becomes removable by equalization [27].

In Fig. 6, additive white Gaussian noise $n(t)$ of SNR=3 [dB] was considered, where the signal-to-ratio noise (SNR) in decibel is defined as $\text{SNR} \triangleq 10 \log_{10} \left( \|x\|^2 / \|n\|^2 \right)$. We fixed $K_B = 10$. Plots (a) and (b) are the results for using the ideal LPF and the random LPF, respectively. Despite the additive noise, the results show that AMWC still reduces the total sampling rate or improves the recovery performance. Including the results in Fig. 6, we conducted more simulations under various SNR=$\{-6, -3, 0, 3, 12\}$ [dB] but omitted to repeat the plots as the graphs exhibit the similar pattern. Instead, we summarized the minimal sampling point results in TABLE III, where the minimal sampling point is defined as the minimal total sampling rate which achieves the support recovery rate of 90%. In the results, as $p$ and/or SNR increase, the minimal sampling point gets smaller, which is expected.

Fig. 7 demonstrates that AMWC reduces the number of channels required for the support recovery. We set $K_B = 10$ and compared the support recovery rates of cMWC and AMWC for various $M$ and given sampling rate of each channel. In plot (a), the support recovery rate of AMWC slightly

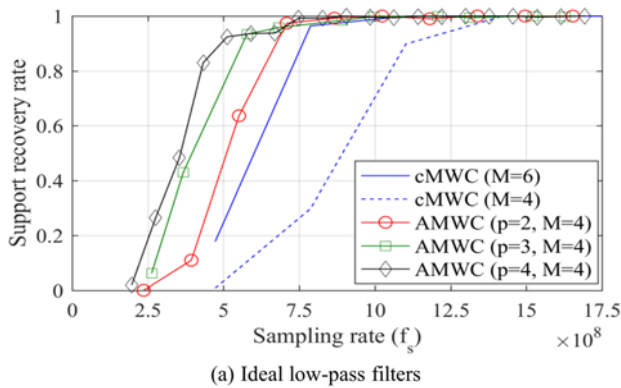

(a) Ideal low-pass filters

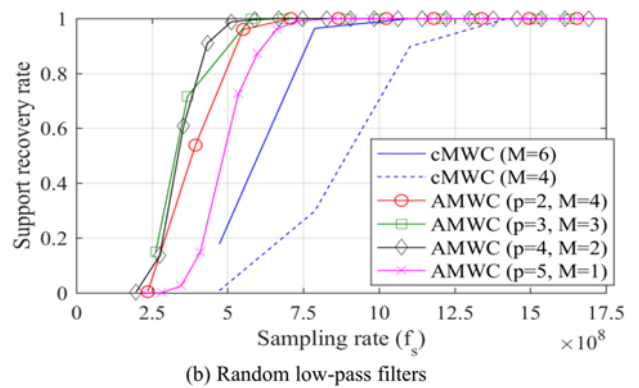

(b) Random low-pass filters

Fig. 7. Rate of successful support recovery of cMWC and AMWC as a function of sampling rate of each channel for various aliasing parameters $p$ and the number of channels $M$. The number of multibands was fixed to $K_B = 10$. Ideal (a) and random (b) low-pass filters were used

outperforms `CMWC`, although `AMWC` uses fewer channels with a lower sampling rate of each channel than `CMWC`. Additionally, in plot (b), when the random low-pass filter is used, `AMWC` using a single channel outperforms `CMWC` using six channels.

As the increase in the number of rows in $\mathbf{Z}$ in (30) or in (43) by $p$-times, the performance of `AMWC` is improved but the computational complexity (CC) for the support recovery with `AMWC` inevitably increases as well. The CC of a compressed sensing algorithm depends on the sizes of matrices in the linear inverse problem $\mathbf{Z}=\mathbf{DX}$. Let $Q_{equation}$, $Q_{sample}$, and $Q_{subband}$ denote the number of rows and columns of $\mathbf{Z}$ and the number of rows of $\mathbf{X}$ for `CMWC` problem, respectively. We make note of the report that the CC of DCS-SOMP with `CMWC` is $O\left(Q_{equation}^2 Q_{subband} Q_{sample}\right)$ [28]. When the two total sampling rates $f_{s,total}$ of `CMWC` and $f'_{s,total}$ of `AMWC` are equal to each other, the number of rows of $\mathbf{Z}$ of `AMWC` becomes $p Q_{equation}$ and that of $\mathbf{X}$ becomes $p Q_{subband}$, respectively, as discussed in Section III-A. In addition, since the bandwidth of the subbands of `AMWC` is $p$-times narrower than that of `CMWC`, the number of columns of $\mathbf{Z}$ becomes $p^{-1}Q_{sample}$. Thus, the CC of DCS-SOMP with `AMWC` is $O\left(p^2 Q_{equation}^2 Q_{subband} Q_{sample}\right)$.

## VI. Conclusion

We proposed a new MWC system called `AMWC` which improves the sampling efficiency by intentionally inducing an aliasing at the ADC. We showed that the improved sampling efficiency leads to reduction on the sampling rate and number of channels required for obtaining a certain number of equations for signal reconstruction. We provided conditions that the sensing matrix of the equations obtained by `AMWC` achieves the Singleton bound, and thus no loss from sampling is guaranteed. In summary, the improved sampling efficiency of `AMWC` reduces the total sampling rate required for lossless sampling. In other words, with fewer channels and less sampling rate of each channel than those of the conventional MWCs, a multiband signal can be captured without information loss by `AMWC`. Conversely, for given hardware resources, the input reconstruction with `AMWC` outperforms the conventional MWCs. Extensive simulation demonstrated that `AMWC` indeed reduces the total sampling rate or improves the reconstruction performance significantly. Additionally, it was demonstrated that the benefits of `AMWC` are maintained in various SNRs. Moreover, use of LPF with random passband response, it was shown, further improves the sampling efficiency.

## Appendix A
## Proof of Lemma 1

With the relationship $f_{LPF} = pf'_s$, the pass-band frequency of $H\left(f - rf'_s\right)$ in (8) is given by $f \in \left[rf'_s - \frac{pf'_s}{2}, rf'_s + \frac{pf'_s}{2}\right]$. When we observe (8) only for a single period $\mathcal{F}'_s(f_0)$, since $W_{LPF} > f'_s$, some of $H\left(f - rf'_s\right)$, the pass bands of which include the frequency domain $\mathcal{F}'_s(f_0)$, can be replaced by the constant frequency response. Without loss of generality, we set the pass-band response to one, i.e., $H(f)=1$ for $f \in \mathcal{F}_{LPF}$. Then, for $r \in \mathbb{Z}$ satisfying

$$rf'_s - \frac{pf'_s}{2} \leq f_0 \tag{44}$$

and

$$rf'_s + \frac{pf'_s}{2} \geq f_0 + f'_s, \tag{45}$$

the shifts of filter responses in (8) are replaced with $H\left(f - rf'_s\right)=1$ within $f \in \mathcal{F}'_s(f_0)$. Let $R_1$ and $R_2$ be the minimum and maximum integers $r$ satisfying (44) and (45), respectively. Additionally, for (8) and (9) to be equivalent, we add some conditions on $R_1$ and $R_2$ such that the pass bands of $H\left(f - rf'_s\right)$ for $r$ smaller than $R_1$ and greater than $R_2$ have no intersection with $f \in \mathcal{F}'_s(f_0)^C$. In other words, we have following conditions on $R_1$ and $R_2$:

$$\left(R_2+1\right)f'_s - \frac{pf'_s}{2} \geq f_0 + f'_s \tag{46}$$

and

$$\left(R_1-1\right)f'_s + \frac{pf'_s}{2} \leq f_0 \tag{47}$$

so that $H\left(f - rf'_s\right)=0$ within $f \in \mathcal{F}'_s(f_0)$ for $r < R_1$ or $r > R_2$. By combining (44) and (46), we have a condition on $R_2$ that

$$R_2 f'_s - \frac{pf'_s}{2} = f_0, \tag{48}$$

and from (45) and (47), we have a condition on $R_1$ that

$$R_1 f'_s + \frac{pf'_s}{2} = f_0 + f'_s. \tag{49}$$

Finally, combining (48) and (49) provides the conditions of Lemma 1. ∎

## Appendix B
## Proofs of Proposition 3 and Lemma 6

### A. Proof of Proposition 3

We track the input-output relation starting from (12):

$$\widetilde{Y}_i\left(e^{j2\pi f T'_s}\right) = \sum_{r=R_1}^{R_2} \sum_{l=-\infty}^{\infty} c_{i,l} X\left(f - (lp+rq)f'_p\right)$$

for $f \in \mathcal{F}'_s(f_0)$, where $R_1$, $R_2$, and $f_0$ satisfy Lemma 1. Alternatively, by using $r' = r - R_1$, we have

$$\begin{aligned}\widetilde{Y}_i\left(e^{j2\pi f T'_s}\right) &= \sum_{r'=0}^{R_2-R_1} \sum_{l=-\infty}^{\infty} c_{i,l} X\left(f - \left(lp+(r'+R_1)q\right)f'_p\right) \\ &= \sum_{r'=0}^{p-1} \sum_{l=-\infty}^{\infty} c_{i,l} X\left(f - \left(lp+(r'+R_1)q\right)f'_p\right)\end{aligned} \tag{50}$$

for $f \in \mathcal{F}'_s(f_0)$, where $R_2 - R_1 = p-1$ by Lemma 1. We replace the term $(r'+R_1)q'$ in (50) by a combination of its quotient $\mu_p(r';q',R_1)$ and remainder $\rho_p(r';q',R_1)$ by divisor $p$, which are, respectively, defined by

$$\mu_p\left(r';q',R_1\right) \triangleq \left\lfloor \frac{(r'+R_1)q'}{p} \right\rfloor \tag{51}$$

and

$$\rho_p\left(r';q',R_1\right) \triangleq \left((r'+R_1)q'\right) \bmod p. \tag{52}$$

By substituting $(r'+R_1)q' = p \cdot \mu_p(r';q',R_1) + \rho_p(r';q',R_1)$ into (50), we have

$$\begin{aligned}
\widetilde{Y}_i\left(e^{j2\pi f T'_s}\right) &= \sum_{l=-\infty}^{\infty}\sum_{r'=0}^{p-1} c_{i,l} X\left(f-\left(lp+p\cdot\mu(r')+\rho(r')\right)f'_p\right) \\
&= \sum_{l=-\infty}^{\infty}\sum_{r'=0}^{p-1} c_{i,l-\mu(r')} X\left(f-\left(lp+\rho(r')\right)f'_p\right)
\end{aligned}$$
(53)

for $f \in \mathcal{F}'_s(f_0)$, where the notations $\mu_p(r';q',R_1)$ and $\rho_p(r';q',R_1)$ are simplified to $\mu(r')$ and $\rho(r')$, respectively. When $p$ and $q'$ are coprime, by modular arithmetic, there exists one-to-one correspondence between $\rho(r')$ and $r'$ modulo $p$. We arrange the order of inner summation of (53) by introducing a utility variable $v \triangleq \rho(r') \in \{0,\cdots,p-1\}$:

$$\begin{aligned}
\widetilde{Y}_i\left(e^{j2\pi f T'_s}\right) &= \sum_{l=-\infty}^{\infty}\sum_{v=0}^{p-1} c_{i,l-\mu\left(\rho_p^{-1}(v;q',R_1)\right)} X\left(f-\left(lp+v\right)f'_p\right)
\end{aligned}$$
(54)

for $f \in \mathcal{F}'_s(f_0)$, where the inverse $\rho_p^{-1}(v;q',R_1)$ of the remainder $\rho_p(r;q',R_1)$ modulo $p$ is computed by

$$\rho_p^{-1}(v;q',R_1) \triangleq \left(v(q')^{-1}-R_1\right)\bmod p, \tag{55}$$

where $(q')^{-1}\bmod p$ is the multiplicative inverse of $q'$ modulo $p$. We simplify the expression $\rho_p^{-1}(v;q',R_1)$ to $\rho^{-1}(v)$. From Lemma 2, we can merge the inner and outer summations of (54) as follows:

$$\widetilde{Y}_i\left(e^{j2\pi f T'_s}\right) = \sum_{k=-\infty}^{\infty} c_{i\left\lfloor\frac{k}{p}\right\rfloor-\mu\left(\rho^{-1}(k\bmod p)\right)} X\left(f-kf'_p\right) \tag{56}$$

for $f \in \mathcal{F}'_s(f_0)$.

We now simplify the picking regularity of the coefficients $c_{i,J(\cdot)}$ in (56), which is defined by

$$\begin{aligned}
J(k;R_1,p,q') &\triangleq \left\lfloor\frac{k}{p}\right\rfloor - \mu\left(\rho^{-1}(k\bmod p)\right) \\
&= \left\lfloor\frac{k}{p}\right\rfloor - \mu\left(\rho^{-1}(k)\right).
\end{aligned}$$
(57)

Meanwhile, by the definitions of the quotient $\mu(\cdot)$ and remainder $\rho(\cdot)$, we have

$$\begin{aligned}
\mu\left(\rho^{-1}(k)\right) &= \left\lfloor\frac{\left(\rho^{-1}(k)+R_1\right)q'}{p}\right\rfloor \\
&= \frac{1}{p}\left(\left(\rho^{-1}(k)+R_1\right)q'-\left(\left(\rho^{-1}(k)+R_1\right)q'\right)\bmod p\right) \\
&= \frac{1}{p}\left(\left(\rho^{-1}(k)+R_1\right)q'-\rho\left(\rho^{-1}(k)\right)\right) \\
&= \frac{1}{p}\left(\left(\rho^{-1}(k)+R_1\right)q'-k\bmod p\right).
\end{aligned}$$
(58)

By substituting (58) into (57),

$$\begin{aligned}
J(k;R_1,p,q') &= \left\lfloor\frac{k}{p}\right\rfloor + \frac{k\bmod p}{p} - \frac{\left(\rho^{-1}(k)+R_1\right)q'}{p} \\
&= \frac{k}{p} - \frac{\left(\rho^{-1}(k)+R_1\right)q'}{p} \\
&= \frac{1}{p}\left\{k-q'\cdot\left[\left(k(q')^{-1}-R_1\right)\bmod p+R_1\right]\right\} \\
&= I(k;R_1,p,q').
\end{aligned}$$
(59)

Thus, the proof is completed. ∎

### B. Proof of Lemma 6

We track the input-output relation starting from (36):

$$\widetilde{Y}_i\left(e^{j2\pi f T'_s}\right) = \sum_{r=R_1}^{R_2}\sum_{l=-\infty}^{\infty} c_{i,l} X\left(f-\left(rq'+lp\right)f'_p\right)G\left(f-rq'f'_p\right)$$

for $f \in \mathcal{F}'_s(f_0)$. Under the conditions of Lemma 1 and Lemma 2, by using $r' \triangleq r-R_1$, we have

$$\begin{aligned}
&\widetilde{Y}_i\left(e^{j2\pi f T'_s}\right) \\
&= \sum_{r'=0}^{R_2-R_1}\sum_{l=-\infty}^{\infty} c_{i,l} X\left(f-\left(lp+(r'+R_1)q'\right)f'_p\right)G\left(f-(r'+R_1)q'f'_p\right) \\
&= \sum_{r'=0}^{p-1}\sum_{l=-\infty}^{\infty} c_{i,l} X\left(f-\left(lp+(r'+R_1)q'\right)f'_p\right)G\left(f-(r'+R_1)q'f'_p\right)
\end{aligned}$$
(60)

for $f \in \mathcal{F}'_s(f_0)$. As done in (50) to (54), we introduce a utility variable $v \triangleq \rho(r')$ and substitute $(r'+R_1)q' = p\cdot\mu\left(\rho^{-1}(v)\right)+v$ into the inputs of $X$ and $G$ in (60). It then follows

$$\widetilde{Y}_i\left(e^{j2\pi f T'_s}\right) = \sum_{l=-\infty}^{\infty}\sum_{v=0}^{p-1}\left(\begin{array}{c} c_{i,J(k;R_1,p,q')} X\left(f-(lp+v)f'_p\right) \\ \cdot G\left(f-\left(p\mu\left(\rho^{-1}(v)\right)+v\right)f'_p\right) \end{array}\right) \tag{61}$$

for $f \in \mathcal{F}'_s(f_0)$. After merging the inner and outer summations based on Lemma 2, we obtain (37)

$$\widetilde{Y}_i\left(e^{j2\pi f T'_s}\right) = \sum_{k=-\infty}^{\infty} d_{i,k}(R_1,p,q') X\left(f-kf'_p\right)G\left(f-\gamma_p(k)f'_p\right)$$

for $f \in \mathcal{F}'_s(f_0)$, where $\gamma_p(k)$ is defined by

$$\begin{aligned}
\gamma_p(k) &\triangleq p\mu\left(\rho^{-1}(k\bmod p)\right)+k\bmod p \\
&= p\mu\left(\rho^{-1}(k)\right)+k\bmod p.
\end{aligned}$$
(62)

By (58) and the definition of $\rho^{-1}(k)$ in (55), (62) turns into

$$\begin{aligned}
\gamma_p(k) &= \left(\rho^{-1}(k)+R_1\right)q' \\
&= q'\left[\left(kq'^{-1}-R_1\right)\bmod p+R_1\right].
\end{aligned}$$
(63)

By the definition of $I(k;R_1,p,q')$ in (15), we finally have

$$\gamma_p(k) = k-pI(k;R_1,p,q'). \tag{64}$$

Thus, the proof is completed. ∎

### APPENDIX C
### PROOF OF PROPOSITION 4

We first show that if $p > q'$ for coprime $p$ and $q'$, at least two columns of $\mathbf{D}$ are identical. Then, from a result in [11], this violates a necessary condition for the unique existence of a $K$-sparse solution.

We first mathematically formulate the meaning of two columns of $\mathbf{D}$ being identical. From Proposition 3 and (30), the entries $d_{i,k+u}(R_1,p,q')$ of $\mathbf{D}$ are picked from $c_{i,J(k;R_1,p,q')}$, where $k$ and $u$ in $d_{i,k+u}$ represent the column and row position, respectively. To search for identical columns in $\mathbf{D}$, we investigate the existence of pairs $(k^*, \omega^*)$ of a column index $k^*$ and shift index $\omega^*$ such that $d_{i,k^*+u} = d_{i,k^*+u+\omega^*}$ for every row index $u \in \mathcal{Q} \triangleq \{0, \cdots, q'-1\}$. In other words, we find pairs $(k^*, \omega^*)$ satisfying

$$I\left(k^*+\omega^*+u; R_1, p, q'\right) = I\left(k^*+u; R_1, p, q'\right). \quad (65)$$

for every $u \in \mathcal{Q}$, where the function $I$ is defined in (15). We use a computation result of $I(k) \triangleq I(k; R_1, p, q')$ in the second line of (59):

$$I(k) = \frac{k}{p} - \frac{\left(\rho^{-1}(k) + R_1\right)q'}{p}, \quad (66)$$

where $\rho^{-1}(k) \triangleq \rho_p^{-1}(k; q', R_1)$ is a function modulo $p$ defined in (55) by $\rho_p^{-1}(k; q', R_1) \triangleq \left(k(q')^{-1} - R_1\right) \bmod p$. By substituting (66) into (65), we rewrite (65) as

$$I\left(k^*+\omega^*+u\right) = I\left(k^*+u\right)$$
$$\Leftrightarrow \rho^{-1}\left(k^*+\omega^*+u\right) = \rho^{-1}\left(k^*+u\right) + \frac{\omega^*}{q'}. \quad (67)$$

We show that, if $p > q'$ and coprime, there exists at least one pair $(k^*, \omega^*)$ of the column index $k^*$ and shifting index $\omega^*$ that satisfy (67) for every row index $u \in \mathcal{Q}$. Before proceeding, we check a computation of $\rho^{-1}\left(k+q'+u\right)$ for every $u \in \mathcal{Q}$. By the definition, it follows

$$\rho^{-1}(k+q'+u) = \left((k+q'+u)(q')^{-1} - R_1\right) \bmod p$$
$$= \left(\left((k+u)(q')^{-1} - R_1\right) \bmod p + 1\right) \bmod p$$
$$= \left(\rho^{-1}(k+u) + 1\right) \bmod p.$$
$$(68)$$

Note that (68) indicates when $\omega^*$ is chosen to $q'$, it satisfies (67), for $k^* \in \mathbb{Z}$ such that $\rho^{-1}\left(k^*+u\right) < p-1$.

What task remains is to show the existence $k^*$ satisfies $\rho^{-1}\left(k^*+u\right) < p-1$ for every row index $u \in \mathcal{Q}$, which implies the existence of identical columns in $\mathbf{D}$ and completes the proof. To this end, we find a set of $\bar{k}\left(\bmod p\right)$ such that $\rho^{-1}\left(\bar{k}+u\right) = p-1$. From the definition, we have

$$\rho^{-1}(\bar{k}+u) \equiv p-1 \ (\bmod p)$$
$$\left(\bar{k}+u\right)(q')^{-1} - R_1 \equiv p-1 \ (\bmod p)$$
$$\bar{k} \equiv (p-1+R_1)q' - u \ (\bmod p)$$
$$\bar{k} \equiv (R_1-1)q' - u \ (\bmod p).$$
$$(69)$$

Note that $(R_1-1)q'$ is a constant. Since the right-hand side of (69) varies by $u \in \mathcal{Q}$, the cardinality of set of $\bar{k}\left(\bmod p\right)$ such that $\rho^{-1}\left(\bar{k}+u\right) = p-1$ is $|\mathcal{Q}| = q'$. Since $p > q'$, this implies there exists $k^*\left(\bmod p\right) \in \{0, 1, \cdots, p-1\}$ such that $\rho^{-1}\left(k^*+u\right) < p-1$, and $k^* \in \mathbb{Z}$ such that $\rho^{-1}\left(k^*+u\right) < p-1$ exists as well.

Consequently, if coprime $p > q'$, there must exist at least one pair of identical columns in $\mathbf{D}$. The existence of identical columns in $\mathbf{D}$ implies $\mathrm{spark}(\mathbf{D}) = 2$. Theorem 2 in [11] states that there exist the unique solution of a linear equation $\mathbf{Z} = \mathbf{D}\mathbf{X}$ for every $K$-sparse solution $\mathbf{X}$ only if

$$K < \frac{\mathrm{spark}(\mathbf{D}) - 1 + \mathrm{rank}(\mathbf{X})}{2}, \quad (70)$$

where $spark$ is the minimum number of linearly dependent columns in $\mathbf{D}$. If $\mathrm{spark}(\mathbf{D}) = 2$, for signals $\mathbf{X}$ with $\mathrm{rank}(\mathbf{X}) \leq 2K-1$, the condition $p > q'$ violates (70). ∎

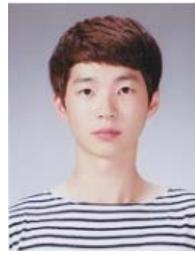
**Jehyuk Jang** received his B.S degree in electronic engineering from the Kumoh National Institute of Technology, Gumi, Korea in 2014 and M.S in Information and Communication Engineering from GIST. Currently, he is pursuing a Ph.D. degree at the School of Electrical Engineering and Computer Science in GIST, Korea. His research interests include sub-Nyquist sampling, compressed sensing.

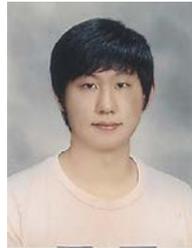
**Sanghun Im** received the B.S. degree in electronics engineering from Soongsil University, Seoul, Korea, in 2009, and the M.S. degree and Ph.D. degree in electrical engineering from Korea Advanced Institute of Science and Technology (KAIST), Daejeon, Korea, in 2011 and 2016, respectively. He is currently working at Hanwha Systems, Korea. His current research interests include communication theories and signal processing for wireless communications and physical layer security.

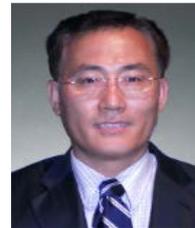
**Heung-No** Lee (SM'13) received the B.S., M.S., and Ph.D. degrees from the University of California, Los Angeles, CA, USA, in 1993, 1994, and 1999, respectively, all in electrical engineering. He then worked at HRL Laboratories, LLC, Malibu, CA, USA, as a Research Staff Member from 1999 to 2002. From 2002 to 2008, he worked as an Assistant Professor at the University of Pittsburgh, PA, USA. In 2009, he then moved to the School of Electrical Engineering and Computer Science, GIST, Korea, where he is currently affiliated. His areas of research include information theory, signal processing theory, communications/networking theory, and their application to wireless communications and networking, compressive sensing, future internet, and brain–computer interface. He has received several prestigious national awards, including the Top 100 National Research and Development Award in 2012, the Top 50 Achievements of Fundamental Researches Award in 2013, and the Science/Engineer of the Month (January 2014).